\documentclass[showpacs,aps,prd,nofootinbib,floatfix,amsmath,amssymb]{revtex4}
\usepackage{amssymb}
\usepackage{amsmath,amssymb,graphicx}
\usepackage{relsize}

\def\lsim{\raise0.3ex\hbox{$\;<$\kern-0.75em\raise-1.1ex\hbox{$\sim\;$}}}
\def\gsim{\raise0.3ex\hbox{$\;>$\kern-0.75em\raise-1.1ex\hbox{$\sim\;$}}}

\begin{document}

\title{$Z'\to ggg$ decay in left-right symmetric models with three and four fermion families}
\author{J. Monta\~no, M. Napsuciale, and C. A. Vaquera-Araujo.}
\affiliation{Departamento de F\'isica, Universidad de Guanajuato,
 Campus Le\'on, Lomas del Bosque 103, Fraccionamiento Lomas del Campestre, C. P. 37150, Le\'on, Guanajuato, M\'exico.}
\begin{abstract}
We study the $Z'\to ggg$ decay in the context of a left-right symmetric gauge theory. We obtain a branching ratio $\mathrm{BR}(Z'\to ggg)=\frac{\Gamma(Z'\to ggg)}
{\Gamma(Z'\to q\bar{q})}= 1.2-2.8 \times 10^{-5}$ for $m_{Z'}=700-1500 ~ GeV$. 
We also study the contribution of a fourth fermion family. We find an enhancement in the branching ratio for $Z'$ masses close to the $\bar{b'}b'$ threshold and a dip for $Z'$ masses close to the $\bar{t'}t'$ threshold. Using the values of the fourth generation quark masses allowed by electroweak precision data we obtain a branching ratio $\mathrm{BR}(Z'\to ggg)= (1-6) \times 10^{-5}$ for $m_{Z'}=(700-1500) ~ GeV$.
\end{abstract}
\maketitle

\section{Introduction}

Additional $Z'$ Gauge Bosons are ubiquitous in Standard Model (SM) extensions. Among them, models based on left-right symmetry groups have been
extensively studied \cite{SMLR-foundations} and are particularly important from the point of view of LHC phenomenology. The basic assumption of left-right symmetric models is that the fundamental weak interaction Lagrangian is invariant under parity symmetry, which is spontaneously broken at low energy due to a non-invariant vacuum. Models based in the smallest left-right symmetric gauge group $SU(3)_C\otimes SU(2)_L\otimes SU(2)_R\otimes U(1)_{B-L}$ have many appealing attributes (for a review see \cite{Mohapatra:1986uf}). These include the same quark-lepton symmetry of the weak interaction and the possibility of writing electric charge in terms of purely physical quantum numbers such as weak-isospin, baryon and lepton number. 

In this work we study the $Z'\to ggg$ rare decay in the context of a Left-Right Symmetric Model based on the $SU(3)_C\otimes SU(2)_R\otimes SU(2)_L\otimes U(1)_{B-L}$ gauge group with three (LRSM) and four fermion families (LRSM4). In general, the coupling $Vggg$ (with $V=Z, Z'$) is absent in the classical action of any renormalizable extension of the SM; however, the process $V\to ggg$ can be induced via quark loops and turns out to be a very interesting prediction which allows us to analyze the interplay between strong and weak sectors of a particular model. The $Vggg$ couplings are also important because they are much less suppressed than those coming from purely weak interactions, like $VVVV$. 

A detailed analysis of the one-loop couplings $Vggg$ and $Vgg\gamma$, with $V=Z,\ Z'$,  in the context of the Minimal 3-3-1 Model \cite{331}, was performed in \cite{V-ggg331,V-gggamma331}. It was explicitly shown there that the $Z\to ggg$ decay \cite{Z-decays} do not receive sizable contribution from quarks in the loops with masses higher than $m_Z/2$ and therefore neither $t$ nor an additional quark family contribute significantly to this process. These results remain valid in LRSM provided the mixing angle between neutral gauge bosons is small which is the case \cite{angle}. 

In LRSM the $Z'\to ggg$ decay is induced by quark loops and the necessary $Z'\bar{q}q$ couplings depend only on a mixing angle which is severely constrained by experimental data. As we shall show below, the most important contributions come from the third family of quarks which motivate us to study also the contributions of a fourth family.

A fourth sequential fermion family is the simplest possible extension to the SM and can be easily adapted to left-right models. It is well-known that the number of families is not fixed by the theory and precision electroweak data do not exclude a fourth one \cite{Maltoni:1999ta, He:2001tp, Novikov:2002tk, Kribs:2007nz, Hung:2007ak, Hashimoto:2010at,Erler:2010sk}. Our results for LRSM suggest that the existence of a fourth generation of quarks could produce an enhancement of the $Z'\to ggg$ branching ratio and it is worthy to quantify this effect. 

An extensive review and an exhaustive list of references to the work on the possible existence of a fourth generation can be found in \cite{Frampton:1999xi}. Recent highlights on the consequences of a fourth 
generation can be found in \cite{Holdom:2009rf}. These include mechanisms of dynamical electroweak symmetry breaking by a fourth generation of quarks and leptons 
\cite{Holdom:1986rn,Hill:1990ge, Carpenter:1989ij, Hung:2009hy, Delepine:2010vw}, convergence improvement of the 
three SM gauge couplings due to the Yukawa coupling contributions from the fourth generation \cite{Hung:1997zj}, 
the possibility of electroweak baryogenesis through first-order electroweak phase transition with four generations 
\cite{Ham:2004xh, Fok:2008yg, Kikukawa:2009mu}, CP violation based on Jarlskog invariants generalized to 
four generations \cite{Hou:2008xd} and the hierarchy problem \cite{Hung:2009ia}. A fourth generation can also solve the CP asymmetry puzzles of $B\to K \pi$ 
\cite{Soni:2008bc,Hou:2005hd,Hou:2006jy} for a range of extra quark masses within the values allowed by  
high precision LEP measurements \cite{Maltoni:1999ta,He:2001tp,Novikov:2002tk}. 

The paper is organized as follows: Section II contains a brief discussion of the LRSM main features, Section III is devoted to the calculation of the $Z'\to ggg$ decay and Section IV summarizes the results. 

\section{The Model}

The LRSM is based in the the Manifest Left-Right Symmetric model developed in \cite{SMLR-main1,SMLR-main2,SMLR-main3-phi-value}, where left and right gauge couplings are equal $g_L=g_R=g$ and Yukawa matrices are hermitian.  In the extended fermion sector, quarks and leptons are placed in doublets with the following assignment of quantum numbers $(d_L,d_R,B-L)$: 
\begin{equation}\nonumber
\begin{array}{ll}
Q_{iL}=\left( \begin{array}{c} u'_i  \\ d_i' \end{array}
\right)_L : (2,1,1/3),
&
Q_{iR}=\left( \begin{array}{c} u'_i  \\ d_i' \end{array}
\right)_R : (1,2,1/3),
\\
 & \\
L_{iL}=\left( \begin{array}{c} \nu'_i  \\ l_i' \end{array}
\right)_L : (2,1,-1),
&
L_{iR}=\left( \begin{array}{c} \nu'_i  \\ l_i' \end{array}
\right)_R : (1,2,-1),
\end{array}
\end{equation}
with $i$=1,2,3. Here $d_{L}$ ($d_{R}$) denotes the dimension of the $SU(2)_L$  ($SU(2)_R$) representation, while the $U(1)$ generator corresponds to $B-L$.  The electric charge formula is given by \cite{B-L}
\begin{equation}\label{}
Q=T_{3L}+T_{3R}+\frac{B-L}{2},
\end{equation}
where $T_{3L}$ and $T_{3R}$ are the $SU(2)_L$ and $SU(2)_R$ generators, respectively. The minimal Higgs sector requires one bidoublet $\Phi:(2,2,0)$ to generate fermion masses \cite{bidoublet}
and two additional triplets $\Delta_{L}:(3,1,2)$ and $\Delta_{R}:(1,3,2)$ to break the $SU(2)_L\otimes SU(2)_R\otimes U(1)_{B-L}$ symmetry down to $U(1)_{\text{em}}$, with the further assumption that only $\Phi$ and $\Delta_R$ have non-vanishing vacuum expectation values (VEVs). 

In this model, after Spontaneous Symmetry Breaking (SSB), we have four non-standard parameters in the gauge sector, {\it i.e.} additional gauge boson masses $M_{W'}$, $M_{Z'}$ and mixing angles in both charged ($\zeta$) and neutral ($\phi$) sectors. The neutral-current Lagrangian in the physical basis can be written as
\begin{equation}\label{neutral}
\mathcal{L}_{NC}^{q} = e\sum_{q=1}^6 Q_{q} (\bar{q}\gamma_\mu q)A^\mu
+\frac{g}{2c_W}\sum_{q=1}^6 [\bar{q}\gamma_\mu(g_{VZ}^{q} 
 -g_{AZ}^{q}\gamma_5)q Z^\mu+\bar{q}\gamma_\mu(g_{VZ'}^{q}-g_{AZ'}^{q}\gamma_5)q {Z'}^\mu] ,
\end{equation}
where $q_1=u,\; q_2=c,\; q_3=t,\; q_4=d,\; q_5=s,\; q_6=b$, with the corresponding weak charges listed in Table \ref{weakcharges}. Here $Z$ is the lightest neutral boson mass eigenstate, while $Z'$ is the heaviest one. The $Z$ boson can be identified with the neutral gauge boson of the SM in the limit of vanishing $\phi$. The current bounds on the mixing angle of the neutral gauge sector $|\phi|\leq 0.0042$ \cite{SMLR-main3-phi-value} yields a prediction for $Z\to ggg$ in LRSM quite similar to that of the Standard Model. 

Adding a fourth generation to this model is straightforward and in Table \ref{weakcharges} we also show the weak charges of the additional quarks which we denote by  $t'$ and $b'$.
\begin{table*}[!ht]
\centering
\caption{\label{weakcharges}  Structure of the neutral currents for the quark sector of the LRSM4.}
\begin{tabular}{cccccc}\hline\hline
Quark & $Q_q$ & $g_{VZ}^q$ & $g_{AZ}^q$ & $g_{VZ'}^q$ & $g_{AZ'}^q$ \\\hline
$u,c,t,t'$ & $+\frac{2}{3}$ & $\frac{3-8s_W^2}{6}\bigg( c_\phi-\frac{s_\phi}{\sqrt{1-2s_W^2}} \bigg)$ & $\frac{c_\phi+s_\phi\sqrt{1-2s_W^2}}{2}$ & $\frac{3-8s_W^2}{6}\bigg( c_\phi+\frac{s_\phi}{\sqrt{1-2s_W^2}} \bigg)$ & $-\frac{c_\phi\sqrt{1-2s_W^2}-s_\phi}{2}$ \\
$d,s,b,b'$ & $-\frac{1}{3}$ & $-\frac{3-4s_W^2}{6}\bigg( c_\phi-\frac{s_\phi}{\sqrt{1-2s_W^2}} \bigg)$ & $-\frac{c_\phi+s_\phi\sqrt{1-2s_W^2}}{2}$ & $-\frac{3-4s_W^2}{6}\bigg( c_\phi+\frac{s_\phi}{\sqrt{1-2s_W^2}} \bigg)$ & $\frac{c_\phi\sqrt{1-2s_W^2}-s_\phi}{2}$ \\\hline\hline
\end{tabular}
\end{table*}

\section{$Z'\to ggg$ Decay}

\begin{figure}
\centering
\includegraphics[width=2in]{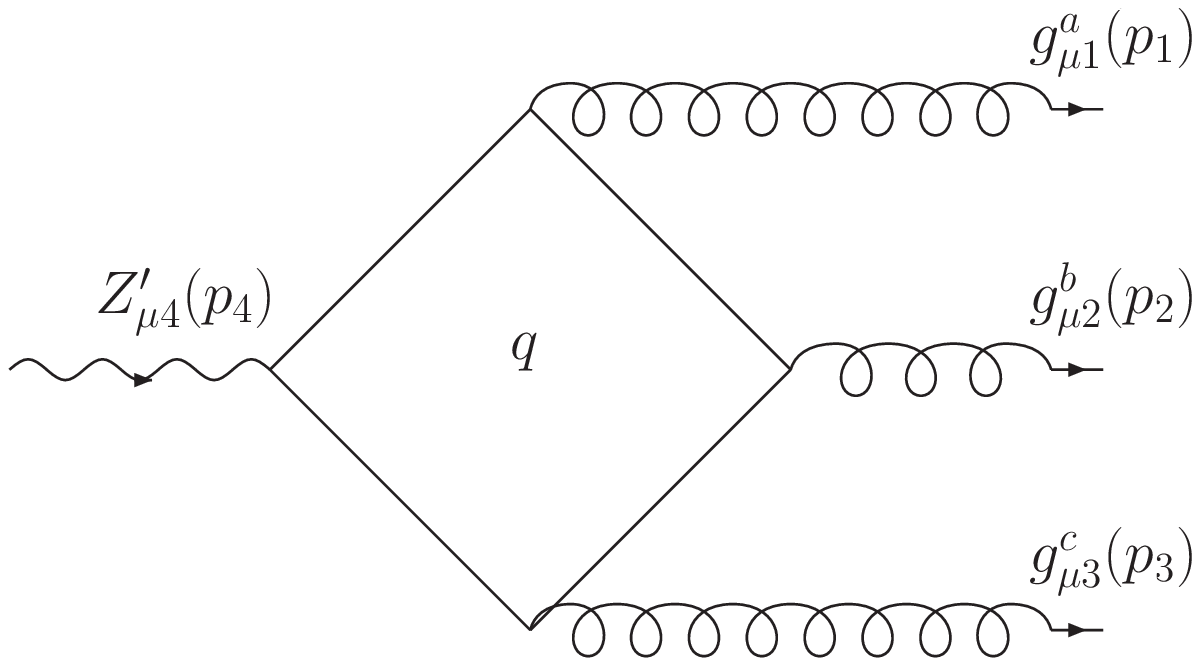} \\
\includegraphics[width=2in]{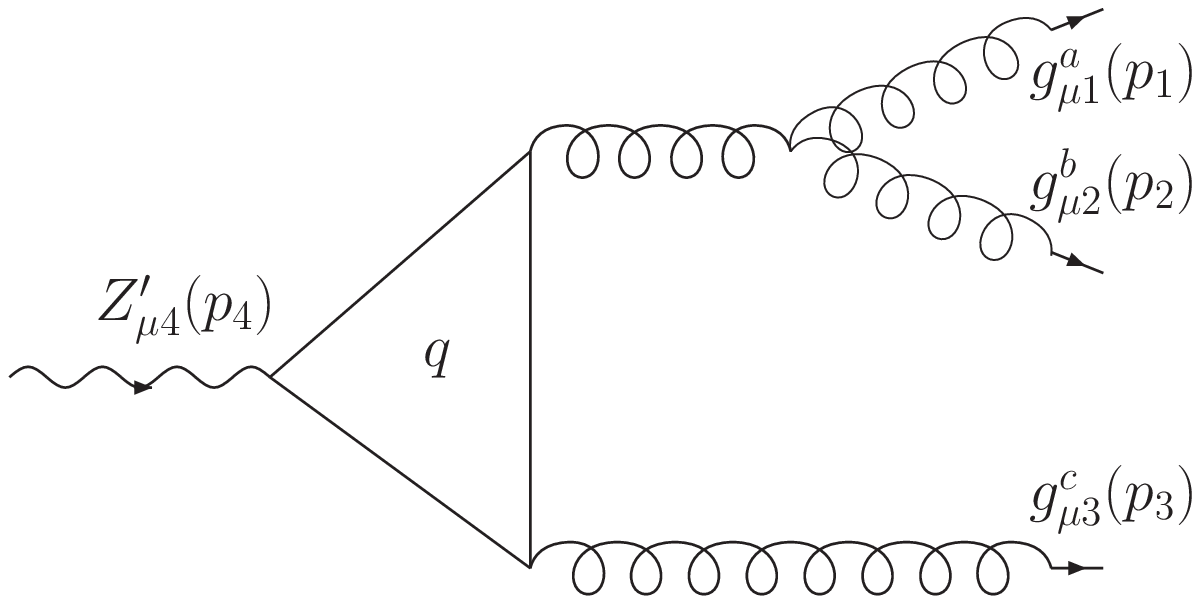} \\
\caption{\label{figure-decays} Examples of box and triangle diagrams contributing to $Z'\to ggg$. }
\end{figure}

In the following we will work in the framework of LRSM4. Results for LRSM can be obtained removing the contributions of the fourth generation quarks.
In LRSM4, the $Z'\to ggg$ decay is induced at one loop level by the box and triangle diagrams depicted in Fig. \ref{figure-decays}. There are six box and six triangle diagrams but one needs to work out only one of each class. Results for the remaining diagrams can be obtained from those of the diagrams in Fig. \ref{figure-decays} using Bose symmetry. 

As discussed in Ref. \cite{V-ggg331}, the invariant amplitude of the process can be written as
\begin{equation}
\mathcal{M}_{Z'\to ggg}=-\frac{ig^3_sg_{Z'}N_C}{4\pi^2}\Big(g^{q}_{VZ'}d_{abc}\mathcal{V}_{q}+
g^{q}_{AZ'}f_{abc}\mathcal{A}_{q}\Big)\;,
\end{equation}
where
\begin{eqnarray}
\mathcal{V}_{q}&=&\sum_{j=1}^{18} f_{V_j}^{q} T_{V_j}^{\mu_1\mu_2\mu_3\mu_4}\epsilon_{\mu_1}^{*\,a}(p_1,\lambda_1) 
  \epsilon_{\mu_2}^{*\,b}(p_2,\lambda_2)
\epsilon_{\mu_3}^{*\,c}(p_3,\lambda_3)\epsilon_{\mu_4}(p_4,\lambda_4),\\
\mathcal{A}_{q}&=&\sum_{j=1}^{24} f_{A_j}^{q} T_{A_j}^{\mu_1\mu_2\mu_3\mu_4}\epsilon_{\mu_1}^{*\,a}(p_1,\lambda_1) 
 \epsilon_{\mu_2}^{*\,b}(p_2,\lambda_2)
\epsilon_{\mu_3}^{*\,c}(p_3,\lambda_3)\epsilon_{\mu_4}(p_4,\lambda_4) \;,
\end{eqnarray}
with $f_{V,A}^q$ as finite form factors of the $T_{V,A}^{\mu_1\mu_2\mu_3\mu_4}$ Lorentz structures, which can be expressed in terms of Passarino-Veltman scalar functions. The corresponding $Z'\to ggg$ decay width is given by
\begin{eqnarray}\label{widthZp-ggg}
  \Gamma(Z'\rightarrow ggg)&=&\frac{m_{Z'}}{3!\;256\; \pi^3}
  \int_0^1\int_{1-x}^1 |\mathcal{M}|^2 dy dx
  =\frac{\alpha_s^3(m_{Z'})\alpha N_C^2 m_{Z'} }{384\;\pi^3c_W^2s_W^2} \times\nonumber \\
&&  \int_0^1\int_{1-x}^1 \sum^{8}_{k,l=1} \bigg[\frac{40}{3}\;g_{VZ'}^{q_{k}}g_{VZ'}^{q_{l}}\bigg(\frac{1}{3}\sum_{\lambda_1,\lambda_2, \lambda_3, \lambda_4}\mathcal{V}_{q_{k}}\mathcal{V}_{q_{l}}^*\bigg)
 +24\;g_{AZ'}^{q_{k}}g_{AZ'}^{q_{l}}\bigg(\frac{1}{3}\sum_{\lambda_1,\lambda_2, \lambda_3, \lambda_4}\mathcal{A}_{q_{k}}\mathcal{A}_{q_{l}}^* \bigg)\bigg] dy dx. 
\end{eqnarray}

The $Z'\to ggg$ decay width can be written as the sum of three partial widths:
\begin{equation}\label{decaywidthZp-ggg}
    \Gamma(Z'\to ggg)=\Gamma_{q}+\Gamma_{q'}+\Gamma_{qq'},
\end{equation}
where $\Gamma_{q}$, $\Gamma_{q'}$, and $\Gamma_{qq'}$ stand for the individual contribution of the SM quarks, the fourth family quarks, $q_{7}=t'$, $q_{8}=b'$, and the interference between both classes, respectively. Passarino-Veltman scalar functions are evaluated numerically using FF routines \cite{FF}. 
For the numerical calculations, we use the values from the Particle Data Group (PDG)  \cite{PDG} for the parameters contained in the amplitude in Eq.~(\ref{widthZp-ggg}):
$ m_u=0_\cdot00255\;\mathrm{GeV}$, $ m_d=0_\cdot00504\;\mathrm{GeV}$, $ m_s=0_\cdot104\;\mathrm{GeV}$, $ m_c=1_\cdot27\;\mathrm{GeV}$, $ m_b=4_\cdot2\;\mathrm{GeV}$, $m_t=171_\cdot2\;\mathrm{GeV}$ and
$s_W^2=0_\cdot23119$.

Regarding the fourth family quark masses, the reported lower bounds from PDG \cite{PDG} are
\begin{equation}\label{}
    m_{t'}>256 \ \mathrm{GeV}, \ \ \ m_{b'}>128 \ \mathrm{GeV},
\end{equation}
but the newest limits on $t'$ and $b'$ masses from direct searches at Tevatron are
$m_{b'} > 338$ GeV \cite{bp-new-mass} and $m_{t'} > 311$ GeV \cite{Lister:2008is}. A rough  upper bound 
$m_{t'} < (4\pi v^2/3)^{1/2}= 504$ GeV can be obtained from unitarity arguments on the s-wave $t'\bar{t}'$ tree level elastic amplitude \cite{Marciano:1989ns}.
According to direct experimental searches for the $Z'$ gauge boson in the context of L-R symmetry, the $Z'$ lower mass limit is $m_{Z'}>630$ GeV at CDF and $m_{Z'}>804$ GeV at LEP 2.  Besides, in Ref.~\cite{angle,Hunt-New-Physics} electroweak precision data requires $m_{Z'}>998$ GeV.

\begin{figure*}
\centering
\includegraphics[width=3.5in]{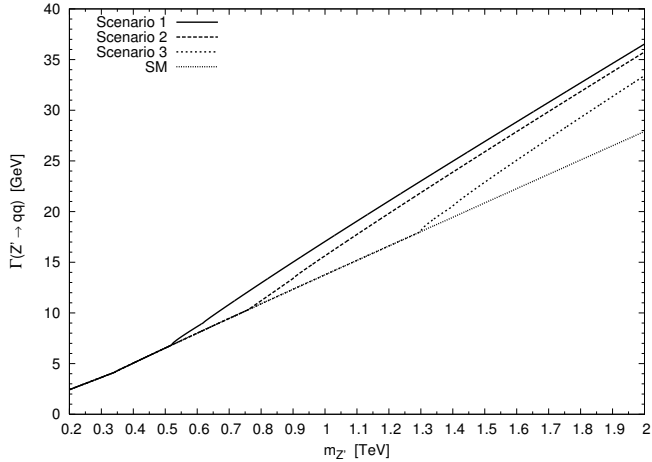}
\caption{ $Z'\to q\bar{q}$ decay width as a function of the $Z'$ mass in LRSM (SM) and in LRSM4 with different values of the fourth generation quark masses:  $\{m_{b'}=260 \ \mathrm{GeV}, \ m_{t'}=310 \ \mathrm{GeV}\}$ (Scenario 1),  $\{m_{b'}=380 \ \mathrm{GeV}, \ m_{t'}=450 \ \mathrm{GeV}\}$ (Scenario 2), and  $\{m_{b'}=650 \ \mathrm{GeV},\ m_{t'}=700 \ \mathrm{GeV}\}$ (Scenario 3).}
\label{figure-Zp-qq_SMLR-Scenarios}  
\end{figure*}

We will present results for the branching ratio 
\begin{equation}\label{}
\mathrm{Br}(Z'\to ggg)=\frac{\Gamma(Z'\to ggg)}{\Gamma(Z'\to q\bar{q})}.
\end{equation}
The width of the leading decay of the $Z'$ boson, $\Gamma(Z'\to q\bar{q})$ is given by 

\begin{equation}\label{Zp-qq}
\Gamma(Z'\rightarrow \bar{q}q)=\frac{\alpha N_Cm_{Z'}}{12 c_W^2s_W^2}\sum_{i=1}^8 f(q_{i}) \;,
\end{equation}
with
\begin{equation}
f(q_{i}) = \sqrt{1-\frac{4m_{q_{i}}^2}{m_{Z'}^2}}
\bigg[ (g_{VZ'}^{q_{i}})^2\left( 1+\frac{2m_{q_{i}}^2}{m_{Z'}^2}\right) 
+(g_{AZ'}^{q_{i}})^2\left( 1-\frac{4m_{q_{i}}^2}{m_{Z'}^2}\right) \bigg]\theta (m_{Z'}^2-4m_{q_{i}}^2),
\end{equation}
with $q_1=u,\; q_2=c,\; q_3=t,\; q_4=d,\; q_5=s,\; q_6=b,\; q_7=t', q_8=b'$. The decay width $\Gamma(Z'\to q\bar{q})$ depends on the unknown $Z'$ mass and on the masses of the fourth generation quarks. In Figure (\ref{figure-Zp-qq_SMLR-Scenarios}) we plot this decay width as a function of the $Z'$ mass in LRSM and using different values of the mass of the fourth generation quarks in the case of LRSM4.

\begin{figure*}
\centering
\includegraphics[width=3.5in]{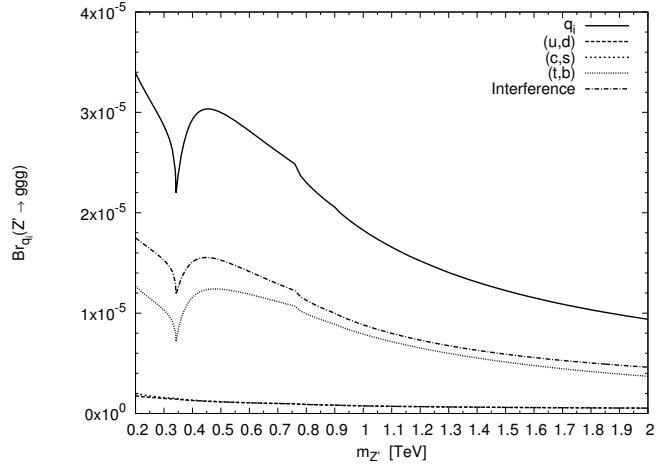}
\caption{\label{LRSMres}  
Separate contribution to BR$(Z'\to ggg)$ for each Standard Model quark family. The line labeled "interference" denotes the cross terms in Eq.(\ref{widthZp-ggg}) when we consider only three fermion families. The solid line yields the total BR$(Z'\to ggg)$ in the LRSM.}
\end{figure*}

\begin{figure*}
\centering
\includegraphics[width=3.5in]{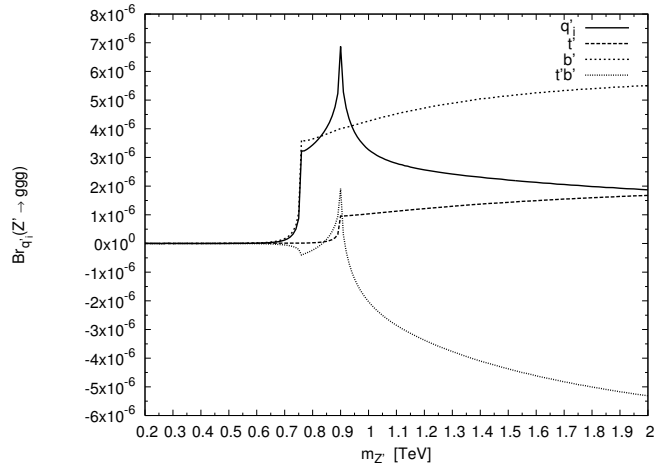}
\caption{\label{fourth} Contributions from the fourth fermion family to BR$(Z'\to ggg)$ as a function of $m_{Z'}$ in LRSM4 for $m_{t'}=450\ \mathrm{GeV},\ m_{b'}=380\ \mathrm{GeV}$: $t'$ (dashed line), $b'$ (short-dashed line), interference (dotted line) and total (solid line).}
\end{figure*}

\begin{figure*}
\centering
\includegraphics[width=3.5in]{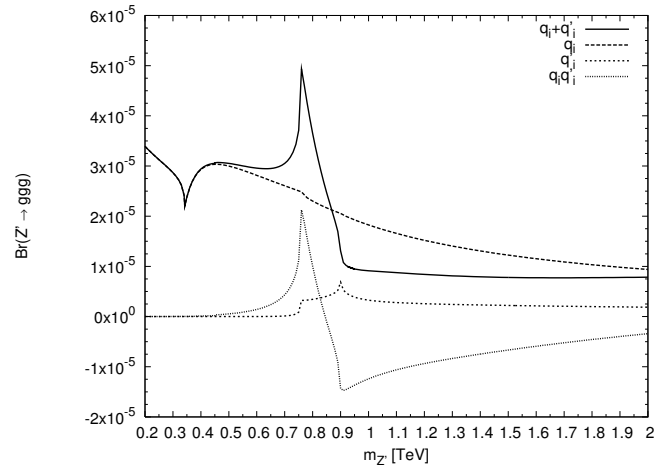}
\caption{\label{figure-scenario3}Contributions to BR$(Z'\to ggg)$ in LRSM4 for $m_{t'}=450$ GeV, $m_{b'}=380$ GeV : Standard Model quarks (dashed), fourth family quarks ( short-dashed), interference (dotted) and total (solid).}
\end{figure*}

\begin{figure*}
\centering
\includegraphics[width=3.5in]{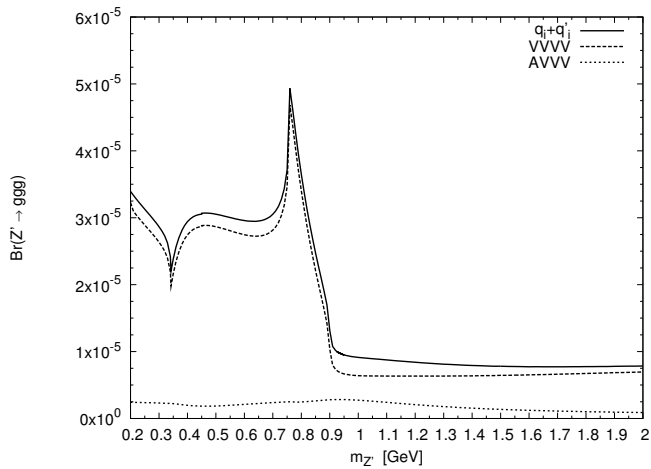}
\caption{\label{figure-scenario1a} Tensor structure of $(Z'\to ggg)$ for $m_{t'}=450\ \mathrm{GeV},\ m_{b'}=380\ \mathrm{GeV}$. Vector VVVV (dashed) and axial vector AVVV (short-dashed) amplitude contributions.}
\end{figure*}

\begin{figure}
\centering
\includegraphics[width=3.5in]{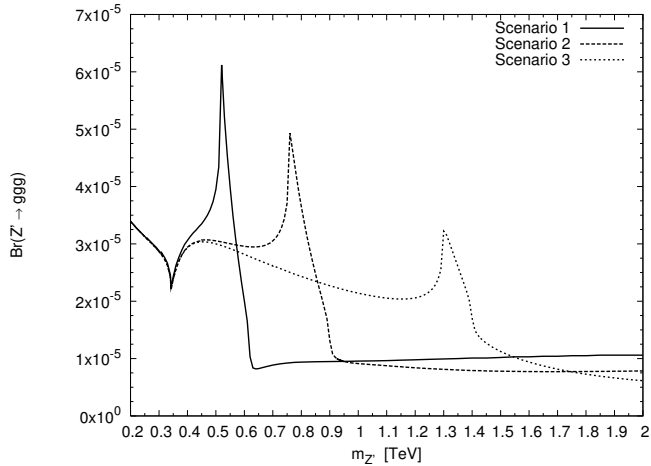}
\caption{\label{figure-scenarios} $BR((Z'\to ggg)$ for different values of the masses of the fourth generation quarks: $\{m_{b'}=260~GeV, ~m_{t'}=310~ GeV \}$ (Scenario 1),  $\{m_{b'}=380~ GeV,~ m_{t'}=450 ~GeV \}$ (Scenario 2), and $\{ m_{b'}=650~ GeV,~ m_{t'}=700 GeV \}$ (Scenario 3) .}
\end{figure}

We present our results for the $\mathrm{Br}(Z'\to ggg)$ as a function of $m_{Z'}$ in Figs. \ref{LRSMres},\ref{fourth},\ref{figure-scenario3},\ref{figure-scenario1a},\ref{figure-scenarios}.  The contributions of the standard model quarks are shown in figure \ref{LRSMres}. The first two families yield roughly the same contributions which are small. In this model, the most important contribution comes from the third family and its interference with the first two families is constructive yielding a total branching ratio of the order $1.2-2.8\times 10^{-5}$ for $700 ~GeV\leq M_{Z'}\leq 1.5~TeV$. Although ruled out by direct searches by CDF and LEP2, in the mass region close to the $\bar{t}t$ threshold the branching ratio abruptly changes due to the opening of this channel in the loops which produces an imaginary part in the amplitude. This opening manifest as a dip in the branching ratio when analyzed as a function of the $Z'$ mass. This rapid change of the branching ratio close to the opening of the $\bar{t}t$ threshold and the fact that the third generation yields the most important contribution makes appealing to study the effects of a fourth family of fermions.
  
We study the effect of a new generation of quarks, considering quark masses below the unitarity constraints and above the lower bound imposed by direct searches. Explicitly, using $m_{t'}=450\ \mathrm{GeV},\ m_{b'}=380\ \mathrm{GeV}$ we obtain the results shown in figure \ref{fourth} from fourth generation quarks in the loops. Notice that the individual contributions are smaller than those of the standard model quarks. Furthermore, in general there is a destructive interference between the $t'$ and $b'$ contributions for the considered masses and this is also true for other values allowed by direct searches and fits to precision electroweak data. However, there is a small region around $m_{Z'}=2m_{t'}$ where this interference is constructive and of the order of the contributions of the third generation, thus an enhancement of the total width for these values of $m_{Z'}$ could be expected but it depends on the relative sign of the fourth generation amplitude with respect to the third one.  The suppression of the individual contributions is natural due to the heavy masses of the fourth generation quarks and the interference pattern comes from the complicated mass dependence of the scalar functions in the loops.

The complete LRSM4 contributions are given in Fig.~\ref{figure-scenario3} for $m_{t'}=450~ GeV, ~m_{b'}=380 ~GeV$.  Notice that the interference between Standard Model quarks and fourth family quarks is constructive for $m_{Z'}\leq 2m_{b'}$ and destructive for $m_{Z'}\geq 2m_{t'}$. In particular, this term reaches its minimal value at $m_{Z'}= 2m_{t'}$  where the fourth generation contribution has its maximal value. This produces an enhancement in the total width at $m_{Z'}=2m_{b'}$ and a  strong dip at $m_{Z'}= 2m_{t'}$. 

It is also interesting to analyze the tensor structure of this decay. The detail of the contributions coming from the axial and vector structures in the invariant amplitude are shown in Fig.~\ref{figure-scenario1a}.  We conclude that the AVVV component is negligible and the decay is predominantly vectorial.

Finally we study the $BR((Z'\to ggg)$ as a function of the masses of the fourth generation quarks. Our results are displayed in Fig. \ref{figure-scenarios} and we can see that the general features discussed above do not depend in detail of the specific values of the fourth generation quark masses.

\section{Summary}

In this work we report a detailed phenomenological analysis of the rare $Z'\to ggg$ decay in a left-right symmetric model with three and four generations. In the conventional L-R model, the predicted branching ratio is in the range $( 1.2-2.8) \times 10^{-5}$ for $m_{Z'}\in 700-1500 ~GeV$ with the most important contribution coming from the quarks in the third family. Furthermore, we show that the vectorial couplings dominate this decay. We also observe an enhancement close to the $\bar{t}t$ threshold which motivate us to study the possible effects of a fourth family of fermions. For masses of the fourth family quarks consistent with precision electroweak data and unitarity bounds we obtain a destructive interference between the up-type ($t'$) and down-type ($b'$) fourth family quarks except for $m_{Z'}$ values around $2m_{t'}$. The complete contributions of the fourth family quarks interfere constructively with those of the Standard Model quarks for $m_{Z'}\approx 2m_{b'}$ and destructively for $m_{Z'}\geq 2m_{t'}$. In the whole the branching ratio for the $Z'\to ggg$ decay lies in the range $(1-6) \times 10^{-5}$ for $m_{Z'}\in 700-1500 ~GeV$.

\acknowledgments{We acknowledge support by CONACYT under project 59741-F, and
SNI (M\' exico). J. Monta\~no would like to thank CONACyT for a posdoctoral fellowship and DCI-UG particle theory group for their hospitality.}

\end{document}